\documentclass[12pt,onecolumn]{IEEEtran}
\IEEEoverridecommandlockouts
\usepackage{epsfig,epstopdf,setspace,subfigure}
\usepackage{amsthm,amsmath,amssymb,cite,enumerate,color}
\newtheorem{proposition}{Proposition}

\newtheorem{remark}{Remark}

\newtheorem{definition}{Definition}

\begin{document}
\title{Transmission Schemes for Gaussian Interference Channels with Transmitter Processing Energy}
\author{Xi~Liu,~\IEEEmembership{Student Member,~IEEE,}
        and~Elza~Erkip,~\IEEEmembership{Fellow,~IEEE}
\thanks{X. Liu is with Broadcom, Matawan, NJ 07747, USA (Email: xi.liu1984@gmail.com).}
\thanks{E. Erkip is with the ECE Department, Polytechnic Institute of New York University, Brooklyn, NY 11201, USA (Email: elza@poly.edu).}
\thanks{This work is presented in part in the Asilomar Conference on Signals, Systems and Computers in Nov. 2011 \cite{processingcost:Liu_Erkip_Asilomar11}.}
}

\maketitle
\doublespace
\begin{abstract}
This work considers communication over Gaussian interference channels with processing energy cost, which explicitly takes into account the energy expended for processing when transmitters are on. In the presence of processing energy cost, transmitting all the time as in the conventional no-cost case is no longer optimal. For a two-user Gaussian interference channel with processing energy cost, assuming that the on-off states of transmitters are not utilized for signaling, several transmission schemes with varying complexities are proposed and their sum-rates are compared with an interference-free upper bound. Moreover, the very strong interference regime, under which interference does not incur any rate penalty, is identified and shown to be larger than the case of no processing energy cost for certain scenarios of interest. Also, extensions to a three-user cascade Gaussian Z interference channel with processing energy cost are provided, where scheduling of user transmissions based on the channel set-up is investigated.
\end{abstract}
\begin{IEEEkeywords}
processing energy, interference channel, bursty transmission.
\end{IEEEkeywords}

\section{Introduction}

In wireless communications, it is often the case that a considerable fraction of the total energy expended by a battery-limited terminal is for processing related to communication. A simple model for this ``processing energy'' is to assume that it is equal to a constant when the transmitter is on. The impact of processing energy for communicating over an additive white Gaussian noise (AWGN) channel was first studied by Youssef-Massaad et al. in \cite{processingcost:Massaad_Medard_Zheng04} \cite{processingcost:Massaad_Medard_Zheng_TWC08}, where it was shown that, under the assumption that turning the transmitter on and off does not convey additional information, Gaussian signaling while keeping the transmitter on for only a fraction of the time is optimal. We will call this strategy ``bursty'' transmission. The authors also extended the analysis to an M-user multiple access channel (MAC) \cite{processingcost:Massaad_Medard_Zheng_Allerton} and showed that time division multiple access outperforms other schemes in terms of the sum rate. In \cite{processingcost:Kramer04}, for the relay channel, Kramer provided a framework for considering processing energy at the source and the relay by modeling power consumed in the sleeping and talking states in a cost function.

\begin{figure}
\centering
\includegraphics[width = 3in]{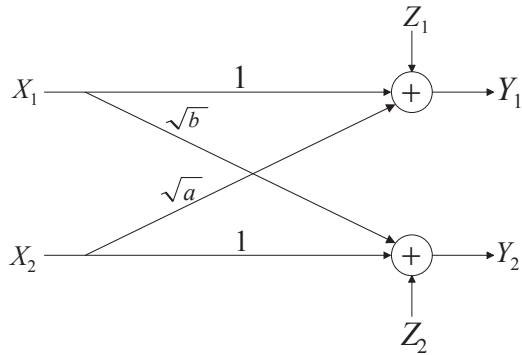}
\caption{Two-user Gaussian interference channel.}
\label{fig:twouserModel}
\end{figure}

\begin{figure}
\centering
\includegraphics[width = 3in]{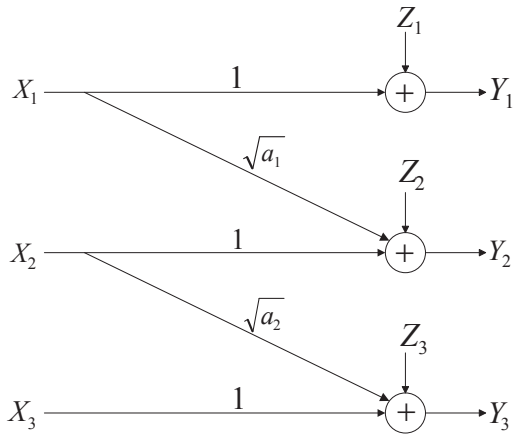}
\caption{Three-user cascade Gaussian Z interference channel.}
\label{fig:cascade}
\end{figure}

This paper extends the previous works by studying the impact of transmitter processing energy cost on interference channels. More specifically, we consider two distinct interference models with transmitter processing energy cost. The first one is a standard two-user Gaussian interference channel (IC) \cite{references:ElGamal_Kim_nit11}, while the second one is a three-user cascade Gaussian Z interference channel (CGZIC) \cite{processingcost:Yuanpeng_Erkip_isit11}. The former, shown in Fig. \ref{fig:twouserModel}, is an information theoretical building block to investigate interference. We assume that the transmitters are energy-limited while the receivers do not have any energy constraints. This could, for example, happen in an up-link scenario in which the transmitters and the receivers are mobile users with limited battery and base stations with stable power supply,  respectively. By studying such a model, we will show how bursty transmission schemes can be employed to mitigate the effect of interference in the presence of processing energy cost. The latter model, shown in Fig. \ref{fig:cascade} on the other hand, is an extension of the two-user Gaussian Z interference channel (ZIC) to the three-user case.  With more than two users in the network, the three-user CGZIC provides the simplest model enabling us to assess the performance advantages of bursty transmission when multiple interfering users need to be scheduled for transmission.

For the case of no processing energy cost, characterization of the capacity region for the general IC remains an open problem; however, the capacity region or the sum-rate capacity for the two-user Gaussian IC is known in special cases such as the strong interference \cite{processingcost:Sato81} \cite{processingcost:Han_Kobayashi81} or the noisy interference \cite{processingcost:Shang09}, and the sum-rate capacity for the three-user CGZIC is known in certain regimes \cite{processingcost:Yuanpeng_Erkip_isit11}. With processing energy cost, the characterization of the capacity region or the sum capacity becomes more involved as each transmitter may choose to be on for only a fraction of the time. Moreover, each transmitter can potentially convey additional information by modulating the on-off states and Gaussian inputs are no longer optimal \cite{processingcost:Kramer04} \cite{processingcost:Zhang_Guo_isit11}. For tractability, as in \cite{processingcost:Massaad_Medard_Zheng_TWC08}, in all the transmission schemes considered in this paper, we will assume the on-off states of each transmitter is known beforehand and thus not utilized for signaling.

For the two-user Gaussian IC with processing energy cost, we discuss several transmission schemes with varying complexities. The sum-rate performances of these schemes are analyzed and compared with an interference-free upper bound. It is shown that, compared with the case of no processing energy cost, in certain cases of interest, a larger range of cross-link power gains ensures interference-free rates, thereby extending the usual very strong interference regime \cite{references:ElGamal_Kim_nit11}. Next, we extend the analysis to the case of three-user CGZIC with processing energy cost to assess the benefits of scheduling users' transmission bursts.

The remainder of this paper is organized as follows. Section II presents the system models and relevant assumptions used. In Section III, for a two-user Gaussian IC with processing energy cost, we propose several transmission schemes with varying complexities to maximize the sum rate and identify the very strong interference regime with processing energy cost. The analysis is then extended to a three-user CGZIC with processing energy cost in Section IV. Finally, we make concluding remarks in Section V.

\section{System Model}
We consider communication for two different types of interference channels with transmitter processing energy cost. The first scenario is a standard two-user two-sided Gaussian IC, as shown in Fig. \ref{fig:twouserModel}, in which two users send messages to their respective receivers, causing interference to each other. The second one is a three-user CGZIC as shown in Fig. \ref{fig:cascade}, in which the first user is interference-free and the second and third users experience only one interference component coming from the first and second users, respectively. This simple three-user interference model is investigated because of its analytical tractability and because of its ability to capture the essence of scheduling over interference channels with more users. For both set-ups, our goal is to design efficient transmission schemes that maximize the achievable sum rate of the system given limited energy budgets at the transmitters.

\subsection{Two-User Gaussian Interference Channel}\label{subsec:modelTwoUser}
The two-user standard Gaussian IC in Fig. \ref{fig:twouserModel} can be expressed as,
\begin{eqnarray}
Y_{1,t} = X_{1,t} + \sqrt{a}X_{2,t}+Z_{1,t},\label{eqn:y1}\\
Y_{2,t} = \sqrt{b}X_{1,t} + X_{2,t} + Z_{2,t},\label{eqn:y2}
\end{eqnarray}
where $X_{i,t}$ and $Y_{i,t}$ represent the input and output of user $i \in \{1,2\}$ at time $t$, respectively, $Z_{1,t}$ and $Z_{2,t}$ are i.i.d. Gaussian noises with zero mean and unit variance, and receiver $i$ is only interested in the message sent by transmitter $i$. Encoding and decoding are done over $n$ channel uses, for $n$ large. For standard definitions of the encoder, the decoder, probability of error and achievable rates, see \cite{references:ElGamal_Kim_nit11}.  Transmitter $i$ is subject to a maximum average power constraint $P_i$. Following \cite{processingcost:Massaad_Medard_Zheng04}, the processing energy cost of transmitter $i$ is modeled as a constant amount $\epsilon_i$ whenever transmitter $i$ is on, with $\epsilon_i\leq P_i$, $i=1,2$. The power constraint at transmitter $i$ is given by
\begin{equation}
\frac{1}{n}\sum_{t=1}^n\left[|X_{i,t}|^2+\epsilon_i \cdot 1_{\{X_{i,t}\neq 0\}}\right]\leq P_i, \label{eqn:avg_power_cnstrnt}
\end{equation}
where $1_{\{\cdot\}}$ is the indicator function.

Due to the processing energy cost, it may not be optimal for each transmitter to transmit all the time. In this case, we can model each transmitter as operating in one of two states: the ``on'' state and the ``off'' state. If transmitter $i$ is in the off state at time $t$, we can model it as transmiting a zero signal, i.e., $X_{i,t}= 0$. If user $j$ ($j\neq i$) is turned off and only user $i$ transmits, by \cite{processingcost:Massaad_Medard_Zheng04}, the optimal transmission scheme for user $i$ when on-off states are fixed (see Section \ref{subsec:OnOffStates} for more discussions on this) is to let transmitter $i$ and receiver $i$ be turned on for a prescribed $\theta_i^*$  fraction of the time with Gaussian signaling of power $\nu_i^*$ such that
\begin{subequations}
\begin{align}
&\theta_i^* = \displaystyle \min\left(1,\frac{P_i W(e^{-1}(\epsilon_i-1))}{(\epsilon_i-1)(W(e^{-1}(\epsilon_i-1))+1)}\right), \label{eqn:thetastar}\\
\text{and}\quad &\nu_i^* = \frac{P_i}{\theta_i^*}-\epsilon_i,\label{eqn:nustar}
\end{align}
\end{subequations}
where $W(\cdot)$ is the LambertW function.\footnote{$W(x)$ is the solution of $We^W = x$.} Note that Eq. (\ref{eqn:thetastar}) above suggests that under processing energy cost, the optimal fraction of time user $i$ should transmit, $\theta_i^*$, generally depends on the average power constraint $P_i$ and the per-channel-use processing energy cost $\epsilon_i$. Moreover, for any given $P_i$, only when $\epsilon_i$ is sufficiently small, we have $\theta_i^* = 1$ as in the case of no processing cost.

If $\theta_1^*+\theta_2^*\leq 1$, it is easily seen that users can employ time division to avoid interference and can thus obtain the same rates as the single user case. In this paper, we primarily focus on the more interesting case $\theta_1^*+\theta_2^*> 1$ when the two users need to compete for the available degrees of freedom.

\subsection{Three-User Cascade Gaussian Z Interference Channel} \label{subsec:modelThreeUser}
The three-user CGZIC as shown in Fig. \ref{fig:cascade} can be expressed as
\begin{eqnarray}
Y_{1,t} = X_{1,t} + Z_{1,t},\label{eqn:cascade_y1}\\
Y_{2,t} = \sqrt{a_1}X_{1,t} + X_{2,t}+Z_{2,t},\label{eqn:cascade_y2}\\
Y_{3,t} = \sqrt{a_2}X_{2,t} + X_{3,t} + Z_{3,t},\label{eqn:cascade_y3}
\end{eqnarray}
where $X_{i,t}$ and $Y_{i,t}$ represent the input and output of user $i \in \{1,2,3\}$ at time $t$, respectively, and $Z_{1,t}$, $Z_{2,t}$ and $Z_{3,t}$ are i.i.d. Gaussian noises with zero mean and unit variance. As in the previous scenario, receiver $i$ is only interested in the message sent by transmitter $i$, and transmitter $i$ is subject to a maximum average power constraint $P_i$ and a constant processing energy cost $\epsilon_i$ joules per time slot when transmitter $i$ is on, with $P_i$ and $\epsilon_i$ satisfying $\epsilon_i\leq P_i$ for $i = 1,2,3$. This leads to (\ref{eqn:avg_power_cnstrnt}) for each transmitter, $i = 1,2,3$. For definitions of the encoder, the decoder, probability of error and achievable rates, the readers are referred to \cite{processingcost:Yuanpeng_Erkip_isit11}.

Moreover, it is assumed that the three users' single-user optimal transmission fractions in (\ref{eqn:thetastar}) satisfy $\theta_1^* + \theta_2^*\geq 1$ and $\theta_2^* + \theta_3^* \geq 1$ to have a non-trivial solution. The transmission schemes discussed later could be modified appropriately to address the cases when either $\theta_1^* + \theta_2^*< 1$ or $\theta_2^* + \theta_3^*< 1$ holds.

\subsection{Fixed Transmitter On-Off States}\label{subsec:OnOffStates}

In our transmission schemes, similar to \cite{processingcost:Massaad_Medard_Zheng04}, we assume on-off states of transmitters are \emph{fixed}, and the receivers are informed beforehand about when each transmitter is in the ``on'' or ``off'' state. Furthermore, we use Gaussian signaling when the transmitters are on.

Note that if on-off states are allowed to be \emph{random} instead, on-off signaling such as pulse position modulation can be employed to transmit additional information to receivers. However, in that case, frequent and fast transition between the two states is required. Accordingly, the energy cost of on-off transition \cite{processingcost:Sinha_Chandrakasan01} cannot be neglected any more and our assumption of modeling the processing energy cost as a constant would break down. Recall that, in the fixed case, each transmitter can remain in the on and off states for long durations of time and very few transitions are needed. Moreover, having the on-off states also carry information may result in non-Gaussian inputs being optimal for $X_1$, $X_2$ and $X_3$, further complicating code design \cite{processingcost:Kramer04}\cite{processingcost:Zhang_Guo_isit11}. Since the on-off states carry at most one bit per channel use, we conjecture that the rate loss due to using fixed transmitter on-off states is at most one bit per user.

\section{Two-User Gaussian IC with Processing Energy Cost}\label{sec:twoUserIC}
For a general two-user Gaussian IC with no transmitter processing energy cost, the best known achievable rate region is given by the full Han-Kobayashi (H-K) rate region \cite{processingcost:Han_Kobayashi81}. The computation of the full H-K region requires taking the
union of all power splits into common and private messages and time sharing, which is difficult due to numerous degrees
of freedom involved \cite{references:Khandani09}. Therefore, for the
purpose of evaluating and computing the achievable sum rate in the processing energy cost case and for practical considerations, we consider several achievable schemes with lower complexity and argue that in certain regimes, the performance is close to optimal. This is in accordance with the no processing energy case, where, for example, a simplified H-K type scheme with fixed power split and no time-sharing is known to achieve a rate region that is within half a bit to the capacity region in \cite{processingcost:Etkin08}.

We now formally define a class of simple H-K schemes, which will be used later for the transmission schemes proposed for the Gaussian IC with processing energy cost.

{\definition \label{def:simpleHK} In a simple H-K scheme for two-user Gaussian IC, user $i$, $i = 1,2$, employs a superimposed Gaussian codebook, where $\tau_i$ portion of the power is used to encode the common information and $(1-\tau_i)$ portion to encode the private information, with $0\leq \tau_i\leq 1$. Receiver $i$ decodes the common part of the interference and its own signal jointly by treating the private part as noise. Hence, each simple H-K scheme in the class is uniquely specified by the power split pair $(\tau_1,\tau_2)$, which we refer to as $\text{HK}(\tau_1,\tau_2)$.}

It follows from \cite{references:Khandani09} that, for the two-user Gaussian IC in Section \ref{subsec:modelTwoUser}, the achievable sum rate for $\text{HK}(\tau_1,\tau_2)$ is given by $\min(\psi_1,\psi_2,\psi_3,\psi_4)$, where
\begin{align}
\psi_{1} &= C\left(\frac{P_1}{1+a(1-\tau_2)P_2}\right) + C\left(\frac{P_2}{1+b(1-\tau_1)P_1}\right),\\
\psi_{2} &= C\left(\frac{P_1+a\tau_2P_2}{1+a(1-\tau_2) P_2}\right) + C\left(\frac{(1-\tau_2)P_2}{1+b(1-\tau_1) P_1}\right),
\end{align}
\begin{equation}
\psi_{3}= C\left(\frac{(1-\tau_1) P_1}{1+a(1-\tau_2) P_2}\right) + C\left(\frac{P_2+b\tau_1P_1}{1+b(1-\tau_1) P_1}\right),
\end{equation}
and
\begin{equation}
\psi_{4} = C\left(\frac{(1-\tau_1) P_1+a\tau_2P_2}{1+a(1-\tau_2) P_2}\right) + C\left(\frac{(1-\tau_2) P_2+b\tau_1P_1}{1+b(1-\tau_1)P_1}\right),
\end{equation}
with $C(x) = (1/2)\log_2(1+x)$. Therefore, the maximum achievable sum rate, maximized over all  $\text{HK}(\tau_1,\tau_2)$ schemes, is given by
\begin{equation}
R_{sum}(P_1,P_2) = \max_{\tau_1,\tau_2}\min(\psi_1,\psi_2,\psi_3,\psi_4).\label{eqn:srate_nobursty}
\end{equation}

{\remark  Under certain conditions, $R_{sum}(P_1,P_2)$ is the sum capacity of the two-user Gaussian IC with $\epsilon_i = 0$ for $i = 1,2$. In the strong interference regime $a\geq 1$ and $b\geq 1$, it is optimal for both the two users to send only common information, i.e., to set $\tau_1 = \tau_2 = 1$ \cite{processingcost:Han_Kobayashi81}; in the noisy interference regime when $a$, $b$, $P_1$ and $P_2$ satisfy $\sqrt{a}(bP_1+1)+\sqrt{b}(aP_2+1)\leq 1$, it is optimal for the users to send only private information, i.e., to set $\tau_1 = \tau_2 = 0$ \cite{processingcost:Shang09}. In general, \cite{processingcost:Etkin08} provides an approximately optimal power split $(\tau_1,\tau_2)$ that performs close to the sum capacity. }


\subsection{Transmission Schemes}\label{subsec:twouserSchemes}
This subsection investigates four different transmission schemes for the general case of processing energy cost ($\epsilon_i>0$, $i = 1,2$). Scheme I uses the $\text{HK}(\tau_1,\tau_2)$ scheme described above without any burstiness, while Scheme II is the TDM scheme when the two users do not overlap their transmission. Different from Scheme I and Scheme II, Scheme III allows for fractional overlap of transmission time and sends independent information over different time fractions. In particular, the $\text{HK}(\tau_1,\tau_2)$ scheme is employed during the overlapped fraction. By contrast, Scheme IV is a generalization of the $\text{HK}(\tau_1,\tau_2)$ scheme under strong interference to the case of processing energy cost, which allows the users to code across different time fractions to achieve higher sum rate.

\subsubsection{Scheme I: Simple H-K Scheme with No Burstiness}
Consider a simple scheme in which both users transmit over all the $n$ time slots. In this case, user $i$ has at most $\nu_i = P_i-\epsilon_i$ joules per time slot for transmission. Using the class of $\text{HK}(\tau_1,\tau_2)$ schemes, the maximum achievable sum rate is
\begin{equation}
R_{sum,I} = R_{sum}(P_1-\epsilon_1,P_2-\epsilon_2),
\end{equation}
where $R_{sum}(\cdot)$ is as in (\ref{eqn:srate_nobursty}).

\subsubsection{Scheme II: Time Division Multiplexing (TDM)}
In this scheme, the two users employ TDM to avoid interference such that user 1 uses $\theta_1$ ($0<\theta_1<1$) fraction of the time while user 2 is left with $1-\theta_1$ fraction. It is easy to see that to maximize the sum rate, it suffices to restrict $\theta_1$ to the range $1-\theta_2^* \leq \theta_1\leq \theta_1^*$. The maximum achievable sum rate can be found by
\begin{equation}
R_{sum,II} = \max_{1-\theta_2^* \leq \theta_1\leq \theta_1^*}\theta_{1}C\left(\frac{P_1}{\theta_1}-\epsilon_1\right)
+(1-\theta_1)C\left(\frac{P_2}{1-\theta_1}-\epsilon_2\right). \label{eqn:sumrate_tdm}
\end{equation}

\subsubsection{Scheme III: Fractional Transmission Overlap, Simple H-K Scheme During the Overlap}
\begin{figure}
\centering
\includegraphics[width = 3in]{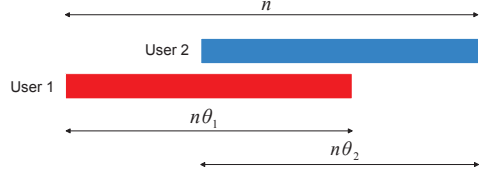}
\caption{Bursty transmission profile $(\theta_1,\theta_2)$ for the two-user Gaussian IC: users 1 and 2 transmit during the first $\theta_1$ fraction and the last $\theta_2$ fraction of the time slots respectively, with $\theta_1+\theta_2-1$ faction of the time slots overlapping.}
\label{fig:fractions}
\end{figure}

Unlike the previous two schemes, Scheme III allows the two users to overlap their transmission over a flexible period of time, as shown in Fig. \ref{fig:fractions}.  Suppose users 1 and 2 transmit over $\theta_1$ and $\theta_2$ fractions of the time respectively, with $\theta_1+\theta_2-1$ fraction overlapping. Since the goal is to maximize the sum rate, the parameters $\theta_1$ and $\theta_2$ need to satisfy
\begin{subequations}
 \begin{align}
  1-\theta_j^* \leq \theta_i &\leq 1,\quad i,j \in\{1,2\}, i\neq j,\\
\theta_1 + \theta_2 &\geq 1.
\end{align}\label{eqn:cnsrtThetasIII}\end{subequations}
We observe that user $i$ does not see any interference for $1-\theta_j$ fraction of the time but suffers from interference for the remaining $\theta_1+\theta_2-1$ fraction, $i, j = 1,2$. We assume that user $i$ sends independent information over these two fractions. For the $(\theta_1+\theta_2-1)$-fraction of overlapping transmissions, the two users employ a given simple H-K scheme, $\text{HK}(\tau_1,\tau_2)$, where the power split pair $(\tau_1,\tau_2)$ can be optimized as in (\ref{eqn:srate_nobursty}). Therefore, the maximum achievable sum rate of Scheme III can be found by
\begin{equation}
R_{sum,III} = \max_{\begin{subarray}{c}\theta_1,\theta_2\end{subarray}}(1-\theta_2)C\left(\frac{P_1}{\theta_1} -\epsilon_1\right) + (1-\theta_1)C\left(\frac{P_2}{\theta_2} -\epsilon_2\right) + (\theta_1+\theta_2-1)R_{sum}\left(\frac{P_1}{\theta_1} -\epsilon_1,\frac{P_2}{\theta_2} -\epsilon_2\right),
\end{equation}
where the maximization is taken over all $\theta_1$ and $\theta_2$ satisfying (\ref{eqn:cnsrtThetasIII}). Note that here we have assumed that user $i$ has the same average power for the $(1-\theta_j)$-fraction and the $(\theta_1+\theta_2-1)$-fraction of its transmission, $j\neq i$. This is motivated by \cite{processingcost:Liu_Erkip_Asilomar11},  where it is numerically shown that for the one-sided interference case ($b = 0$), allowing power control brings little performance gain over the no power control case.

\subsubsection{Scheme IV: Fractional Transmission Overlap, Joint Encoding/Decoding}
In Scheme III, we assume the two users send independent information over different fractions of the time. However, coding across each user's entire transmission can be advantageous, since each receiver may leverage the information overheard during the fraction of the time when its own transmitter is off to facilitate decoding of the interference signal. Note that this scheme is most appealing when interference is strong, i.e., $a\geq 1$ and $b\geq 1$, since the overheard signal is stronger than the signal received at the intended receiver. Thus, in Scheme IV, we assume both users send only common messages and each receiver jointly decodes its own information and the interfering user's information using the signals received during the entire $n$ time slots.

In order to obtain the maximum achievable sum rate of Scheme IV for $a\geq 1$ and $b\geq 1$, we first obtain an achievable rate region by having each receiver jointly decode both users' messages.  We divide the total $n$ time slots into three fractions as shown in Fig. \ref{fig:fractions}, and assume that user $i$ transmits with constant power $P_i/\theta_i-\epsilon_i$ during the $\theta_i$ fraction of the time when its transmitter is on. With each receiver decoding both users' messages, the Gaussian IC becomes equivalent to a Gaussian compound MAC. For any given choice of $(\theta_1,\theta_2)$, using standard arguments \cite{references:ElGamal_Kim_nit11}, if both transmitters use Gaussian codebooks, the rate pairs $(R_1,R_2)$ in the following rate region can be shown to be achievable
\begin{subequations}
\begin{align}
&\hspace{3cm}R_1\leq \theta_{1} C\left(\frac{P_1}{\theta_1}-\epsilon_1\right),\label{eqn:sumRate_nonred1}\\
&\hspace{3cm}R_2\leq \theta_{2} C\left(\frac{P_2}{\theta_2}-\epsilon_2\right),\\
R_1 + R_2\leq& \min\left\{(\theta_1+\theta_2-1)C\left(\frac{P_1}{\theta_1}-\epsilon_1+a\left(\frac{P_2}{\theta_2}-\epsilon_2\right)\right)+ (1 - \theta_{2}) C\left(\frac{P_1}{\theta_1}-\epsilon_1\right)\right.\nonumber\\
&+ (1-\theta_{1})C\left(a\left(\frac{P_2}{\theta_2}-\epsilon_2\right)\right),(\theta_1+\theta_2-1)C\left(b\left(\frac{P_1}{\theta_1}-\epsilon_1\right)+\frac{P_2}{\theta_2}-\epsilon_2\right)\nonumber\\
 &\left.+ (1 - \theta_{2}) C\left(b\left(\frac{P_1}{\theta_1}-\epsilon_1\right)\right) + (1-\theta_{1})C\left(\frac{P_2}{\theta_2}-\epsilon_2\right)\right\}.\label{eqn:sumRate_red}
\end{align}\end{subequations}
Taking the union over all $(\theta_1,\theta_2)$ satisfying (\ref{eqn:cnsrtThetasIII}), we obtain the desired achievable rate region for $a\geq 1$ and $b\geq 1$. Then, the maximum achievable sum rate by Scheme IV can be obtained immediately as
\begin{align}
R_{sum,IV} =  &\max_{\theta_1,\theta_2} \min\left\{\theta_{1}C\left(\frac{P_1}{\theta_1} -\epsilon_1\right) + \theta_{2}C\left(\frac{P_2}{\theta_2} -\epsilon_2\right)\right.,\nonumber\\
 &\hspace{-0.1cm}(\theta_1+\theta_2-1)C\left(\frac{P_1}{\theta_1} -\epsilon_1+a\left(\frac{P_2}{\theta_2} -\epsilon_2\right)\right)+(1-\theta_2)C\left(\frac{P_1}{\theta_1} -\epsilon_1\right)+(1-\theta_1)C\left(a\left(\frac{P_2}{\theta_2} -\epsilon_2\right)\right),\nonumber\\
&\hspace{-0.2cm}\left.(\theta_1+\theta_2-1)C\left(b\left(\frac{P_1}{\theta_1} -\epsilon_1\right)+\frac{P_2}{\theta_2} -\epsilon_2\right)+(1-\theta_2)C\left(b\left(\frac{P_1}{\theta_1} -\epsilon_1\right)\right)+(1-\theta_1)C\left(\frac{P_2}{\theta_2} -\epsilon_2\right) \right\},\label{eqn:srate4}
\end{align}
where the maximization is taken over all $(\theta_1,\theta_2)$ satisfying (\ref{eqn:cnsrtThetasIII}).

We remark that, Scheme IV is a generalization of the simple H-K scheme from the conventional case of continuous transmission to the processing energy case when both users employ a bursty transmission profile as in Fig. \ref{fig:fractions}, for the strong interference regime $a\geq 1$ and $b\geq 1$. It is generally superior to Scheme III in terms of the achievable sum rate, since coding the two users' messages across all the $n$ time slots improves communication rates compared to independent encoding and decoding for different fractions. However, in Scheme IV, receiver $i$ needs to be on even when transmitter $i$ is silent.

\subsection{Very Strong Interference Regime with Processing Energy Cost}
In this subsection, we determine the range of power gains $(a,b)$ for which interference does not incur any rate penalty to either user. The no-loss range is referred to as \textit{very strong interference regime with processing energy cost}. Recall that, in the case of no processing energy cost, the very strong interference regime is given by $a\geq 1+P_1$ and $b\geq 1+P_2$. Our main contribution in this subsection is the derivation of a new very strong interference regime when processing energy costs are taken into account.

{\proposition \label{prop:verystrong}The two users in the Gaussian IC with processing energy cost can both achieve their maximum interference-free rates $(C(\nu_1^*),C(\nu_2^*))$ if the following conditions are satisfied:
\begin{subequations}
\begin{align}
1+\nu_2^*\leq& \left(1+a\nu_2^*\right)^{\rho_1} \left(1+\frac{a\nu_2^*}{1+\nu_1^*}\right)^{1-\rho_1},\label{eqn:condition1_1}\\
1+\nu_1^*\leq& \left(1+b\nu_1^*\right)^{\rho_2} \left(1+\frac{b\nu_1^*}{1+\nu_2^*}\right)^{1-\rho_2},\label{eqn:condition1_2}
\end{align}\label{eqn:condition1_12}\end{subequations}
where $\rho_1 = (1-\theta_1^*)/\theta_2^*$ and $\rho_2 = (1-\theta_2^*)/\theta_1^*$. Here $\theta_i^*$ and $\nu_i^*$ represent user $i$'s optimal burstiness and signal power level in the interference-free case given in (\ref{eqn:thetastar}) and (\ref{eqn:nustar}) respectively.
}
\begin{proof}
Using the achievable rate region bounded by (\ref{eqn:sumRate_nonred1})-(\ref{eqn:sumRate_red}), in order for both users to achieve their interference-free rates as in the single user case, the inequality in (\ref{eqn:sumRate_red}) should be redundant for transmission fractions $(\theta_1^*,\theta_2^*)$ as in (\ref{eqn:thetastar}). Mathematically, this is satisfied if
\begin{subequations}
\begin{align}
\theta_1^*C(\nu_1^*) \leq& (1-\theta_2^*)C(b \nu_1^*)+(\theta_1^*+\theta_2^* - 1)C\left(\frac{b\nu_1^*}{1+\nu_2^*}\right), \label{eqn:condition1}\\
\theta_2^*C(\nu_2^*) \leq& (1-\theta_1^*)C(a \nu_2^*)+(\theta_1^*+\theta_2^* - 1)C\left(\frac{a\nu_2^*}{1+\nu_1^*}\right), \label{eqn:condition2}
\end{align}\end{subequations}
where $\nu_i^* = P_i/\theta_i^* - \epsilon_i$. These conditions can be simplified as (\ref{eqn:condition1_12}) in Proposition \ref{prop:verystrong} since $\rho_1 = (1-\theta_1^*)/\theta_2^*$ and $\rho_2 = (1-\theta_2^*)/\theta_1^*$.
\end{proof}

The regime in Proposition \ref{prop:verystrong} can be considered as a generalization of the very strong interference regime to the processing overhead case. Note that in the case of no processing overhead, we have $\rho_i = 0$ and the conditions in (\ref{eqn:condition1_12}) reduce to the usual very strong interference regime.

To further simply these conditions is difficult in general. In the following, we focus on the special case when $\epsilon_i\rightarrow 0$, $P_i\rightarrow 0$ and $P_i/ \sqrt{2\epsilon_i} = \lambda_i$ for some constant $\lambda_i>0$, $i = 1,2$. In this case, $\theta_i^*$ in (\ref{eqn:thetastar}) can be simplified as $\theta_i^*=\min(1,\lambda_i)>0$. Hence $\rho_i\geq 0$. Also, by (\ref{eqn:nustar}) it follows that $\nu_i^* \rightarrow 0$. Using Taylor series approximation, we have
\begin{subequations}
\begin{align}
1+\nu_2^*\leq& (1+\rho_1 a\nu_2^*) \left(1+(1-\rho_1)\frac{a\nu_2^*}{1+\nu_1^*}\right),\label{eqn:condition3_1}\\
\text{and}\quad 1+\nu_1^*\leq& (1+\rho_2 b\nu_1^*) \left(1+(1-\rho_2)\frac{b\nu_1^*}{1+\nu_2^*}\right).\label{eqn:condition3_2}
\end{align}
\end{subequations}
Ignoring terms containing $(\nu_1^*)^2$ or $(\nu_2^*)^2$ on the right side of (\ref{eqn:condition3_1}) and (\ref{eqn:condition3_2}), we find the following sufficient conditions
\begin{equation}
a\geq \frac{1+\nu_1^*}{1+\rho_1\nu_1^*}, \quad \text{and}\quad b\geq \frac{1+\nu_2^*}{1+\rho_2\nu_2^*}.\label{eqn:condition2_12}
\end{equation}
Moreover, given $\rho_1\geq 1-\theta_1^*$, we have
\begin{subequations}
\begin{align}
\frac{1+\nu_1^*}{1+\rho_1\nu_1^*}&\leq \frac{1+\nu_1^*}{1+(1-\theta_1^*)\nu_1^*} \\
&= \frac{(1+\nu_1^*)(1+\theta_1^*\nu_1^*)}{1+\nu_1^* + \theta_1^*(1-\theta_1^*)(\nu_1^*)^2}\\
&\leq 1+\theta_1^*\nu_1^*\\
&< 1+ \theta_1^*(\nu_1^*+\epsilon_1)= 1+P_1. \label{eqn:largervstr}
\end{align}\end{subequations}
Similarly, we have $(1+\nu_2^*)/(1+\rho_2\nu_2^*) < 1+P_2$. Thus, the conditions in (\ref{eqn:condition2_12}) suggest that a larger range of power gains $a$ and $b$ ensure very strong interference under processing energy cost.

Depending on the values of $\lambda_1$ and $\lambda_2$, we can write out the exact very strong interference regime as in the following:
 \begin{itemize}
 \item $\lambda_i<1$ ($i=1,2$): in this case, $\nu_i^* \thickapprox \sqrt{2\epsilon_i}$ and $\theta_i^* = \lambda_i$. Accordingly, $\rho_1 =(1-\lambda_1)/\lambda_2$ and $\rho_2 =(1-\lambda_2)/\lambda_1$. Thus, we can simply the inequalities in (\ref{eqn:condition2_12}) as
\begin{subequations}
\begin{align}
a\geq &\frac{P_2+\sqrt{2\epsilon_1}P_2}{P_2+\sqrt{2\epsilon_2}(\sqrt{2\epsilon_1}-P_1)}\triangleq \bar{a},\\
b\geq &\frac{P_1+\sqrt{2\epsilon_2}P_1}{P_1+\sqrt{2\epsilon_1}(\sqrt{2\epsilon_2}-P_2)}\triangleq \bar{b}.
\end{align}\end{subequations}
Note that $\theta_1^*+\theta_2^* = \frac{P_1}{\sqrt{2\epsilon_1}}+\frac{P_2}{\sqrt{2\epsilon_2}}>1$, i.e., $\sqrt{2\epsilon_1}P_2 >\sqrt{2\epsilon_2}(\sqrt{2\epsilon_1}-P_1)$; hence, $\bar{a}>1$. Moreover, we have proved that $\bar{a}< 1+P_1$ in (\ref{eqn:largervstr}). Thus, we have $1<\bar{a}< 1+P_1$. Similarly, $1<\bar{b}< 1+P_2$.
\item $\lambda_1<1$, $\lambda_2\geq 1$: in this case, $\nu_1^* \thickapprox \sqrt{2\epsilon_1}$, $\theta_1 = \lambda_1$, $\theta_2^* = 1$ and $\nu_2^* = P_2-\epsilon_2$. Accordingly, $\rho_1 = 1 - \lambda_1$ and $\rho_2 = 0$. Thus, the inequalities in (\ref{eqn:condition2_12}) can be simplified as $a \geq \frac{1+\sqrt{2\epsilon_1}}{1+\sqrt{2\epsilon_1}-P_1}$ and $b\geq 1+P_2-\epsilon_2$.
\item $\lambda_1\geq 1$, $\lambda_2<1$: similar to the case $\lambda_1<1$, $\lambda_2\geq 1$, we can simplify the inequalities in (\ref{eqn:condition2_12}) as $a\geq 1+P_1-\epsilon_1$ and $b \geq \frac{1+\sqrt{2\epsilon_2}}{1+\sqrt{2\epsilon_2}-P_2}$.

\item $\lambda_i\geq 1$ ($i = 1,2$): in this case, $\theta_1^* = \theta_2^* = 1$ and hence $\rho_1 =\rho_2 = 0$. The inequalities in (\ref{eqn:condition2_12}) degenerates into the one in the trivial case $a\geq 1+P_1$ and $b\geq 1+P_2$.
\end{itemize}

\subsection{Illustration of Results}\label{subsec:twouserResults}
\begin{figure}
\centering
\includegraphics[width = 3.5in]{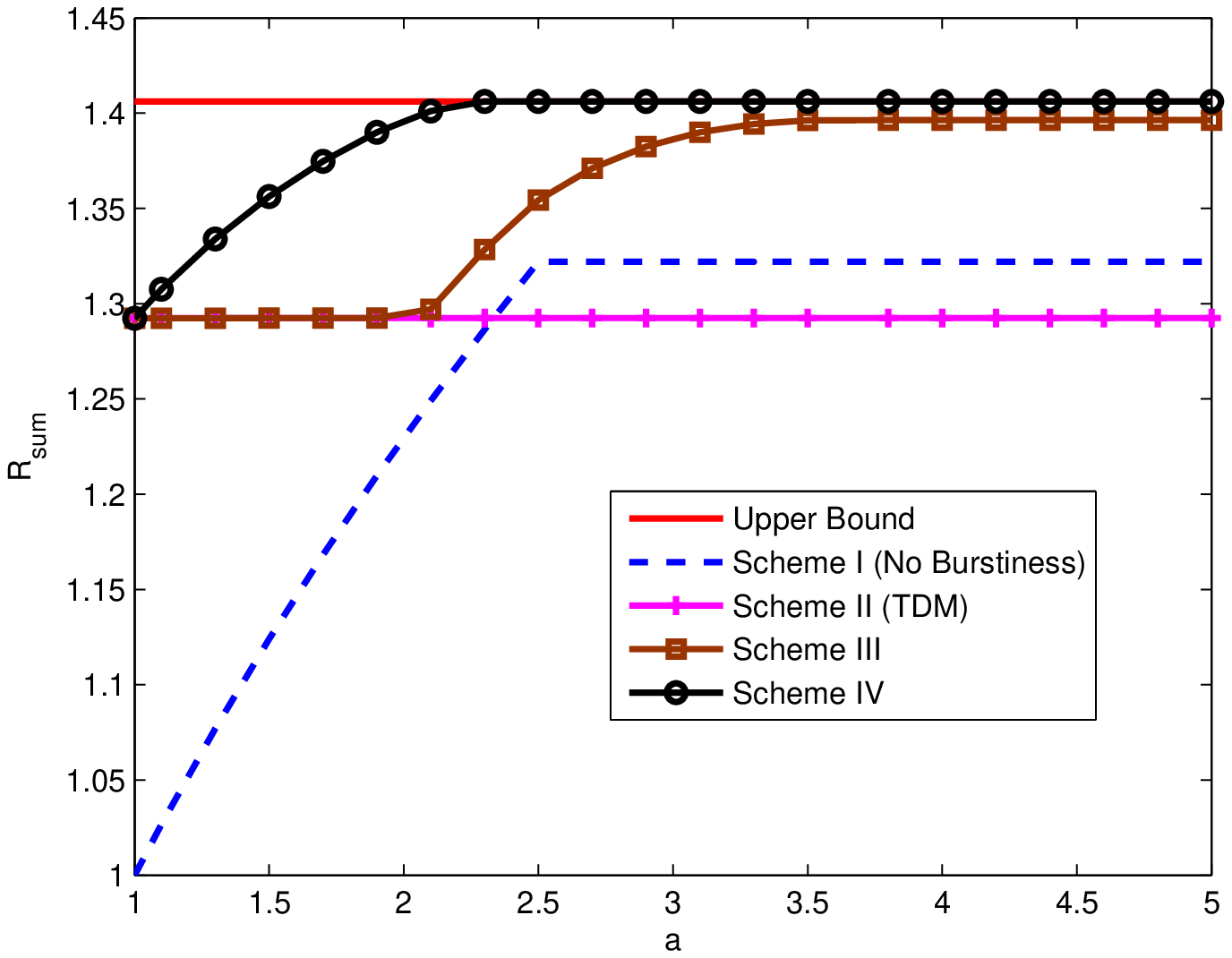}
\caption{The maximum sum rates achieved by various schemes for the two-user Gaussian IC, $R_{sum}$, along with the interference-free upper bound, as a function of cross-link power gain $a$ for $a\geq 1$, when $b = 3$, $P_1 = P_2 = 3.5$ and $\epsilon_1 = \epsilon_2 = 2$.}
\label{fig:fig2a}
\end{figure}

\begin{figure}
\centering
\includegraphics[width = 3.5in]{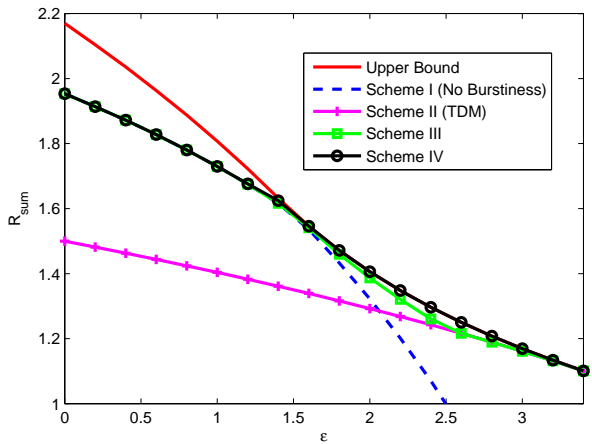}
\caption{The maximum sum rates achieved by various schemes for the two-user Gaussian IC, $R_{sum}$, along with the interference-free upper bound, as a function of processing energy cost $\epsilon$, under the assumption that $\epsilon_1 = \epsilon_2 = \epsilon$, when $a = b = 3$ and $P_1 = P_2 = 3.5$.}
\label{fig:srateVSepsilon}
\end{figure}

\begin{figure}
\centering
\includegraphics[width = 3.5in]{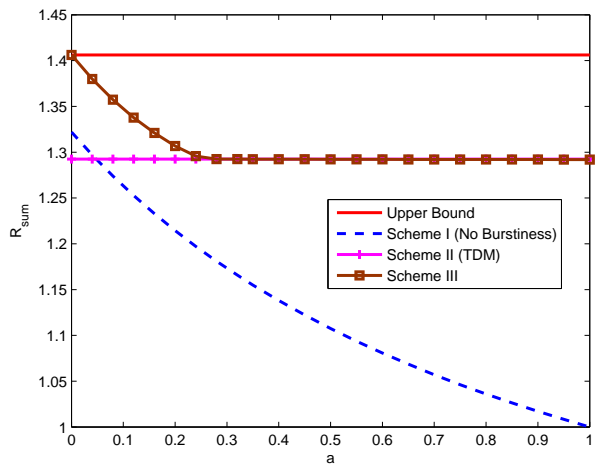}
\caption{The maximum sum rates achieved by various schemes for the two-user Gaussian ZIC, $R_{sum}$, along with the interference-free upper bound, as a function of cross-link power gain $a$ for $0<a<1$, when $b = 0$, $P_1 = P_2 = 3.5$ and $\epsilon_1 = \epsilon_2 = 2$.}
\label{fig:fig4}
\end{figure}

In this subsection, we numerically evaluate the achievable sum rate of the proposed transmission schemes for the two-user Gaussian IC through examples. Fig. \ref{fig:fig2a} plots the sum rates as a function of cross-channel power gain $a$ for $a\geq 1$ when $b = 3$, $P_1 = P_2 = 3.5$, $\epsilon_1 = \epsilon_2 = 2$. Since $a\geq 1$ and $b\geq 1$, the optimal power split $(\tau_1,\tau_2)$ that maximizes the sum rate in (\ref{eqn:srate_nobursty}) is given by $(1,1)$ \cite{processingcost:Sato81} \cite{processingcost:Han_Kobayashi81}. Given these parameters, we have $\theta_1^* = \theta_2^* = 0.76$ and $\nu_1^* = \nu_2^* = 2.59$ \cite{processingcost:Massaad_Medard_Zheng04}, and thus the condition $\theta_1^*+\theta_2^*>1$ is satisfied. For comparison, the sum of users' maximum interference-free rates is provided as an upper bound. In Fig. \ref{fig:fig2a}, Scheme IV outperforms the others in terms of sum rate. Moreover, it achieves the upper bound when $a$ is greater than 2.3. This is consistent with Proposition \ref{prop:verystrong}, which says that in this case, the very strong interference regime corresponds to $a\geq 2.3$ and $b\geq 2.3$. As TDM does not depend on $a$ and $b$, the sum rate of Scheme II remains constant. Depending on the value of $a$, Scheme I and Scheme II may outperform one another. By allowing a flexible overlap of time slots, Scheme III can generally gain better sum rates than Scheme I and Scheme II. It is worth noting that, the sum rate achieved by TDM is strictly less than that of Scheme IV except when $a = 1$. In contrast, recall that in \cite{processingcost:Massaad_Medard_Zheng_Allerton}, under the same assumption of fixed on-off states, TDM was shown to be the best scheme in maximizing the sum rate for the MAC with processing energy cost.

In Fig. \ref{fig:srateVSepsilon}, assuming $\epsilon_1 = \epsilon_2 = \epsilon$, we plot the sum rates as a function of $\epsilon$ when we set $ a = b = 3$ and $P_1 = P_2 = 3.5$. We observe that, for $\epsilon\leq 1.2$, Scheme I, Scheme III and Scheme IV have the same performance, which is strictly superior to Scheme II. This is consistent with the intuition that non-bursty transmission remains optimal for sufficiently small processing energy costs. As $\epsilon$ gradually increases, the performance of Scheme I deteriorates fast, Scheme IV starts to outperform Scheme III and meets the interference-free upper bound at $\epsilon = 1.6$. At the same time, Scheme II results in a higher sum rate than Scheme I as $\epsilon$ grows larger than 2.1, and becomes equivalent to Scheme III when $\epsilon$ increases to 2.6. Finally, Scheme II coincides with the interference-free upper bound and Scheme IV when $\epsilon$ reaches 3.4. This is because, for $\epsilon\geq 3.4$, due to the very large processing energy cost, each user's optimal burst fraction in the single-user case is smaller than 0.5, and therefore TDM can be employed to avoid interference.

Next, we evaluate the sum rate performance of the proposed schemes for the special case of Gaussian ZIC with weak interference, i.e., $0< a< 1$ and $b = 0$. The result will serve as a basis for the analysis of the sum rate of the three-user CGZIC in Section IV. Fig. \ref{fig:fig4} plots the sum rate as a function of cross-link power gain $a$ for $0<a<1$ when $P_1 = P_2 = 3.5$ and $\epsilon_1 = \epsilon_2 = 2$ as in Fig. \ref{fig:fig2a}. Scheme IV, which is tailored for $a\geq 1$ and $b\geq 1$, is not considered. Since $b = 0$, the optimal power split $(\tau_1,\tau_2)$ that maximizes the sum rate in (\ref{eqn:srate_nobursty}) is given by $(0,0)$ \cite{processingcost:Shang09}. As shown in Fig. \ref{fig:fig4}, Scheme III has the best sum-rate among all three schemes and is strictly better than TDM for $a\leq 0.28$, which suggests that allowing the two users to overlap their transmission and treating interference as noise during the overlap is beneficial when interference is sufficiently weak. However, as $a$ grows beyond 0.28, TDM starts to coincide with Scheme III in terms of the sum rate. This implies that TDM is in fact the best scheme in terms of maximizing the sum rate for a large range of moderately weak $a$'s.

\begin{figure}
\centering
\includegraphics[width = 3.5in]{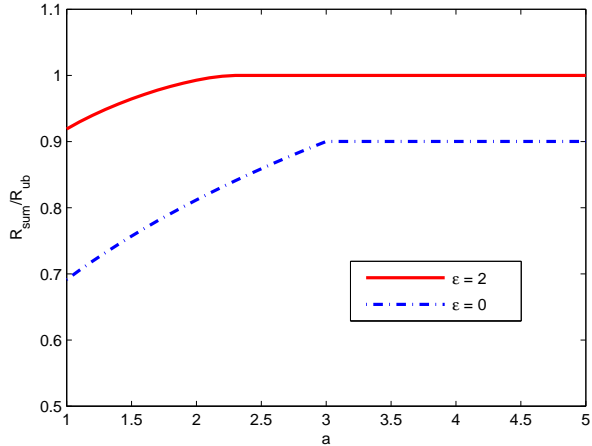}
\caption{Normalized maximum achievable sum rates $R_{sum}/R_{ub}$ for the two-user Gaussian IC with processing energy cost $\epsilon_1 = \epsilon_2 = \epsilon$, as a function of cross-link power gain $a$ for $\epsilon = 2$ and $\epsilon = 0$ when $b = 3$, and $P_1 = P_2 = 3.5$, where $R_{sum} = R_{sum,IV}$ and $R_{ub}$ is the interference-free upper bound.}
\label{fig:ICRatio}
\end{figure}

Finally, in order to quantitatively evaluate how the destructive effect of interference can be mitigated via bursty transmission in the case of processing energy cost, in Fig. \ref{fig:ICRatio}, assuming $\epsilon_1 = \epsilon_2 = \epsilon$, we compare the maximum achievable rate $R_{sum}$, normalized by the interference-free upper bound $R_{ub}$, i.e., the ratio $R_{sum}/R_{ub}$, for the processing energy case $\epsilon = 2$ and the no processing energy case $\epsilon = 0$. Since $a\geq 1$ and $b\geq 1$, $R_{sum}$ is set to $R_{sum,IV}$ in (\ref{eqn:srate4}), which reduces to (\ref{eqn:srate_nobursty}) for the case of no processing energy cost. It can be observed from Fig. \ref{fig:ICRatio} that, for all $a\geq 1$, the normalized sum rate in the case of $\epsilon = 2$ is substantially larger than its counterpart in the case of $\epsilon = 0$. This demonstrates that leveraging bursty transmission as in Scheme IV reduces the rate loss incurred by interference substantially in the processing energy cost case. Moreover, in the case of $\epsilon = 2$, the normalized sum rate reaches 1 for sufficiently large $a$ while in the case of $\epsilon = 0$, it saturates at 0.9. This is because the very strong interference regime with no processing energy cost requires $b\geq 1+P_2 = 4.5$ but we have $b=3$ in this example.

\section{Three-User CGZIC with Processing Energy Cost}
This section extends the analysis on the impact of transmitter processing energy cost in interference channels from the two-user Gaussian IC to the three-user cascade CGZIC introduced in Section \ref{subsec:modelThreeUser}. Similar to Section \ref{subsec:twouserSchemes}, we study several achievable schemes with relatively low complexity for the purpose of maximizing the sum rate. As mentioned in Section \ref{subsec:modelThreeUser}, we focus on the non-trivial case $\theta_1^* + \theta_2^*\geq 1$ and $\theta_2^* + \theta_3^* \geq 1$. Moreover, we study the mixed interference regime $a_1\geq 1$ and $0<a_2<1$ in detail, for which we show how the optimal transmission bursts of all the users are interconnected. The other regimes for $a_1$ and $a_2$ will also be discussed at the end of this section.

We note that, the class of simple H-K schemes defined in Definition \ref{def:simpleHK} can be extended in a straightforward manner to the three-user CGZIC. That is,
user $i$, $i = 1,2,3$, employs a superimposed Gaussian codebook, with $\tau_i$ portion of the power used to encode the common information and $(1-\tau_i)$ portion used to encode the private information, where $0\leq \tau_i\leq 1$. Receiver $i$ decodes the common part of the interference and its own signal jointly by treating the private part as noise. With a total of three users, each simple H-K scheme is then uniquely specified by the power split tuple $(\tau_1,\tau_2,\tau_3)$, which we refer to as $\text{HK}(\tau_1,\tau_2,\tau_3)$. Moreover, in \cite{processingcost:Yuanpeng_Erkip_isit11}, Liu and Erkip argued that, with no processing energy cost, for all the power gains $(a_1,a_2)$, setting $\tau_i = 0\text{ or }1$, $i=1,2,3$, is optimal under the class of simple H-K schemes described above, leading to a maximum sum rate
\begin{equation}
R_{sum}(P_1,P_2,P_3) = \sum_{i=1}^3 C(\gamma_iP_i),\label{eqn:cascZ_3user}
\end{equation}
where $\gamma_1 = 1$ and for $i = 2,3$,
\begin{equation}
\gamma_i = \begin{cases}
\frac{1}{1+a_{i-1}P_{i-1}},\; a_{i-1}\leq \gamma_{i-1}\\
\min\left(\frac{(a_{i-1}-\gamma_{i-1})P_{i-1}+P_i}{P_i+\gamma_{i-1}P_{i-1}P_i},1\right),\; a_{i-1}>\gamma_{i-1}
\end{cases}.\label{eqn:gamma_i}
\end{equation}
For $a_1\geq 1$ and $0<a_2<1$, \cite{processingcost:Yuanpeng_Erkip_isit11} shows that it is optimal to let user 1 send only common information (i.e., $\tau_1 = 1$), to let user 2 send only private information (i.e., $\tau_2 =0$) if $a_2\leq \gamma_2$ and only common information (i.e., $\tau_2 = 1$) if $a_2>\gamma_2$, and to let user 3 send only private information (i.e., $\tau_3 = 0$). Under certain conditions, $R_{sum}(P_1,P_2,P_3)$ is known to be either equal to or close to the sum capacity of the three-user CGZIC. For example, for a mixed regime with $1\leq a_1< 1+P_2$ and $a_2\leq 1/(1+a_1P_1)$, $R_{sum}(P_1,P_2,P_3)$ is within half a bit to the sum capacity. The readers are referred to \cite{processingcost:Yuanpeng_Erkip_isit11} for more details.

In the following, using the above results, we provide several communication schemes for the three-user CGZIC in the case of processing energy cost. These schemes are similar to those studied in Section \ref{subsec:twouserSchemes}. The emphasis here will be how one can schedule the users'  transmission bursts based on the fact that there is at most one interference component at each receiver in the CGZIC set-up. For convenience, notations similar to those in Section \ref{subsec:twouserSchemes} are used.

\subsection{Transmission Schemes}
\subsubsection{Scheme I: Simple H-K Scheme with No Burstiness}
Consider a simple scheme in which all the three users transmit over all the $n$ time slots, i.e., $\theta_1=\theta_2=\theta_3 = 1$. In this case, user $i$ has at most $\nu_i = P_i-\epsilon_i$ joules per time slot for transmission. Using the class of $\text{HK}(\tau_1,\tau_2,\tau_3)$ schemes, the maximum achievable sum rate is given by \begin{equation}
R_{sum,I} = R_{sum}(P_1-\epsilon_1,P_2-\epsilon_2,P_3-\epsilon_3),
\end{equation}
where $R_{sum}(\cdot)$ is as in (\ref{eqn:cascZ_3user}).

\subsubsection{Scheme II: TDM}
In this scheme, users employ TDM to avoid interference. Note that, since users 1 and 3 do not interfere with each other in the cascade Z set-up, we allow their transmission to overlap as much as possible, but impose that user 2's transmission does not interfere with that of either user 1 or user 3. That is, if users 1 and 3 transmit $\theta_1$ and $\theta_3$ fractions of the time, then user 2 is left with the remaining $1-\max(\theta_1,\theta_3)$ fraction, where $1-\theta_2^*\leq \theta_1\leq \theta_1^*$ and $1-\theta_2^*\leq \theta_3\leq \theta_3^*$. The maximum achievable sum rate of TDM can be found by
\begin{equation}
R_{sum,II} = \max_{\theta_1,\theta_3}\theta_{1}C\left(\frac{P_1}{\theta_1}-\epsilon_1\right)+\theta_{3}C\left(\frac{P_3}{\theta_3}-\epsilon_3\right)
+(1-\max(\theta_1,\theta_3))C\left(\frac{P_2}{1-\max(\theta_1,\theta_3)}-\epsilon_2\right).
\end{equation}

\subsubsection{Scheme III: Fractional Transmission Overlap, Simple H-K Scheme During the Overlap}
\begin{figure}
\centering
\includegraphics[width = 3in]{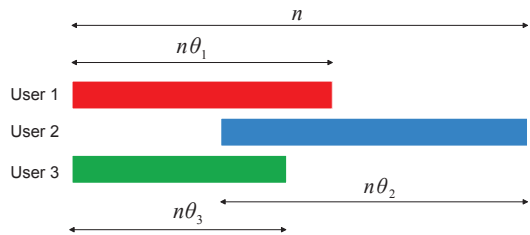}
\caption{Bursty transmission profile $(\theta_1,\theta_2,\theta_3)$ for the three-user CGZIC: users 1 and 3 transmit during the first $\theta_1$ and $\theta_3$ fractions of the time respectively, while user 2 transmits during the last $\theta_2$ fraction of the time.}
\label{fig:cascadeFractions}
\end{figure}
For the set-up in Fig. \ref{fig:cascade}, if we consider the ZIC with the first two users only, from \cite{processingcost:Liu_Erkip_Asilomar11}, it follows that it is advantageous to have user 2's transmission overlap with user 1's for a certain fraction since $a_1>1$, i.e., $\theta_2>1-\theta_1$. On the other hand, for the ZIC with the last two users only, for $0<a_2<1$, the numerical results in Section \ref{subsec:twouserResults} indicate that in most cases it is better to let users 2 and 3 operate in a time-division manner, i.e., users 2 and 3's transmission bursts should satisfy $\theta_2 = 1-\theta_3$. This suggests there is generally a trade-off in selecting $\theta_2$. Given the above observations, in this scheme, user 2 may overlap its transmission partially with users 1 and 3. As in Scheme II, since users 1 and 3 do not interfere with each other, they maximize the transmission overlap. Hence, without loss of optimality, we assume that the three users transmit using the bursty transmission profile $(\theta_1,\theta_2, \theta_3)$ as shown in Fig. \ref{fig:cascadeFractions}, where users 1 and 3 transmit during the first $\theta_1$ and $\theta_3$ fractions of the time respectively, while user 2 transmits during the last $\theta_2$ fraction. For simplicity, the parameters $(\theta_1,\theta_2,\theta_3)$ are assumed to satisfy\footnote{More generally, we may allow user 2 to operate in a TDM manner with one of user 1 and user 3, i.e., $\theta_1 + \theta_2 < 1$ or $\theta_2 + \theta_3 < 1$, which is not investigated here for the sake of simplicity.}
\begin{subequations}
\begin{align}
&1-\theta_2^*\leq \theta_k \leq 1,\quad k = 1,3,\\
&1-\max(\theta_1^*,\theta_3^*)\leq \theta_2\leq 1,\\
\text{and }\;&\theta_2+\min(\theta_1,\theta_3)\geq 1.
\end{align}\label{eqn:bursty_profile}\end{subequations}
Moreover, as in Scheme III of Section \ref{sec:twoUserIC}, we assume each user sends independent information over different fractions of the time with constant power, i.e., we have $\nu_i = P_i/\theta_i-\epsilon_i$, $i = 1,2,3$. For any transmission profile $(\theta_1,\theta_2,\theta_3)$ with $\theta_1 \geq \theta_3$, the resulting sum rate for all the users is given by
\begin{align}
R_{sum,III}(\theta_1,\theta_2,\theta_3) = &(1-\theta_1)C\left(\frac{P_2}{\theta_2}-\epsilon_2\right)+(1-\theta_2)\left(C\left(\frac{P_1}{\theta_1}-\epsilon_1\right)+C\left(\frac{P_3}{\theta_3}-\epsilon_3\right)\right) \nonumber\\ &+(\theta_2+\theta_3-1)R_{sum}\left(\frac{P_1}{\theta_1}-\epsilon_1,\frac{P_2}{\theta_2}-\epsilon_2,\frac{P_3}{\theta_3}-\epsilon_3\right)+(\theta_1-\theta_3)\nonumber\\ &\cdot\min\left\{C\left(\frac{P_1}{\theta_1}-\epsilon_1\right)+C\left(\frac{P_2}{\theta_2}-\epsilon_2\right),C\left(a_1\left(\frac{P_1}{\theta_1}-\epsilon_1\right)+\frac{P_2}{\theta_2}-\epsilon_2\right)\right\}.\label{eqn:sumRateIII}
\end{align}
Note that in (\ref{eqn:sumRateIII}), for the $(\theta_2+\theta_3-1)$-fraction when all the users transmit, we have the three-user sum rate $R_{sum}(P_1/\theta_1-\epsilon_1,P_2/\theta_2-\epsilon_2,P_3/\theta_3-\epsilon_3)$ of (\ref{eqn:cascZ_3user}), while for the $(\theta_1-\theta_3)$-fraction when only users 1 and 2 transmit, we can use the strong interference sum rate for the two-user Gaussian ZIC \cite{processingcost:Sason_IT04}. Similarly, we can obtain the sum rate for any transmission profile $(\theta_1,\theta_2,\theta_3)$ with $\theta_1<\theta_3$. Finally, the maximum achievable sum rate of Scheme III, $R_{sum,III}$, can then be obtained through optimizing over all the transmission profiles $(\theta_1,\theta_2, \theta_3)$ satisfying (\ref{eqn:bursty_profile}).

\subsubsection{Scheme IV: Fractional Transmission Overlap, Successive Interference Cancelation at Receiver 2}
In this scheme, we assume the three users still transmit using the bursty profile $(\theta_1,\theta_2, \theta_3)$ as shown in Fig. \ref{fig:cascadeFractions}, with $(\theta_1,\theta_2, \theta_3)$ constrained to satisfy (\ref{eqn:bursty_profile}), and with constant signal powers. However, since $a_1>1$, similar to Scheme IV of Section III, we allow user 2's receiver to listen to user 1's transmission when its own transmitter is off to facilitate decoding of the interference from user 1. We will further require that user 2 perfectly cancels the interference from user 1. This is possible if user 1 transmits with a rate no larger than\footnote{Note that user 1's rate needs to chosen to satisfy two constraints: 1) Receiver 1 can successfully decode its signal; 2) Receiver 2 can successfully decode user 1's signal by treating its own signal as noise.}
\begin{equation}
R_1 =  \min\left\{\theta_1 C\left(\frac{P_1}{\theta_1}-\epsilon_1\right),(1-\theta_2)C\left(a_1\left(\frac{P_1}{\theta_1}-\epsilon_1\right)\right)+(\theta_1+\theta_2-1)C\left(\frac{a_1\left(P_1/\theta_1-\epsilon_1\right)}{1+P_2/\theta_2-\epsilon_2}\right)\right\}.\label{eqn:burJointRate1}
\end{equation}
After interference cancelation, user 2 sees an interference-free link. Therefore, the three-user CGZIC effectively decomposes into a two-user Gaussian ZIC with users 2 and 3 being the transmitters, and a separate point-to-point link for user 1 transmitting at rate $R_1$. Since $0<a_2<1$, treating interference as noise during the overlap is optimal, the maximum sum rate of users 2 and 3 can be written as
\begin{equation}
R_2+R_3 = \theta_2 C\left(\frac{P_2}{\theta_2}-\epsilon_2\right) + (1-\theta_2)C\left(\frac{P_3}{\theta_3}-\epsilon_3\right) + (\theta_2+\theta_3-1)C\left(\frac{P_3/\theta_3-\epsilon_3}{1+a_2\left(P_2/\theta_2-\epsilon_2\right)}\right).\label{eqn:burJointRate23}
\end{equation}
The achievable sum rate for any given transmission profile $(\theta_1,\theta_2, \theta_3)$ is obtained by summing (\ref{eqn:burJointRate1}) and (\ref{eqn:burJointRate23}). Finally, the maximum achievable sum rate of Scheme IV, $R_{sum,IV}$, can be found by optimizing over all the transmission profiles $(\theta_1,\theta_2,\theta_3)$ satisfying (\ref{eqn:bursty_profile}).

\subsection{Illustration of Results}
\begin{figure}
\centering
\includegraphics[width = 3.5in]{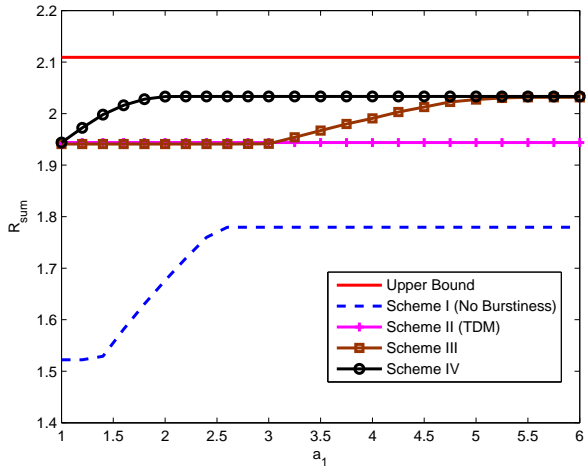}
\caption{The maximum sum rates achieved by various schemes for the three-user CGZIC, along with the interference-free upper bound, as a function of cross-link power gain $a_1$ when $a_2 = 0.5$, $P_1 = 4$, $P_2 = 3.5$, $P_3 = 3$ and $\epsilon_1 = \epsilon_2 = \epsilon_3 = 2$.}
\label{fig:cascRate}
\end{figure}
\begin{figure}
\centering
\includegraphics[width = 3.5in]{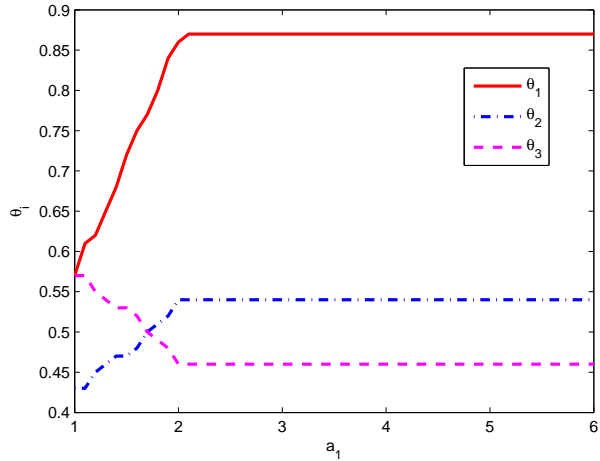}
\caption{The optimal transmission fractions of the three users in Scheme IV as a function of cross-link power gain $a_1$ when $a_2 = 0.5$,  $P_1 = 4$, $P_2 = 3.5$, $P_3 = 3$ and $\epsilon_1 = \epsilon_2 = \epsilon_3 = 2$.}
\label{fig:cascBursti}
\end{figure}

Fig. \ref{fig:cascRate} compares the maximum achievable sum rates of various schemes as a function of cross-link power gain $a_1$ for a mixed regime example when $a_1\geq 1$ and $a_2 = 0.5$, with $P_1 = 4$, $P_2 = 3.5$, $P_3 = 3$, and $\epsilon_1 = \epsilon_2 = \epsilon_3 = 2$. The interference-free upper bound is also plotted for comparison. If there is no interference, the optimal single-user transmission fractions for the three users are $\theta_1^* = 0.87$, $\theta_2^* = 0.76$ and $\theta_3^* = 0.65$, respectively. It is seen in Fig. \ref{fig:cascRate} that, Scheme I has the worst sum rate among all. We also observe that, the curve for Scheme III coincides with that for TDM for $a_1$ smaller than 3. This implies that when $a_1$ is not sufficiently large, there is no benefit for user 2 to overlap its transmission with either user 1 or user 3. However, as $a_1$ grows larger, Scheme III starts to dominate TDM, since users 1 and 2 can gain by allowing fractional transmission overlap. Scheme IV is superior to the other schemes except for very large $a_1$'s when Scheme III performs the same as Scheme IV. This shows the importance of interference overhearing and cancelation at receiver 2 in Scheme IV.

The optimal transmission fractions of the three users in Scheme IV are plotted in Fig. \ref{fig:cascBursti} for the same parameters as in Fig. \ref{fig:cascRate}. It can be observed that, when $a_1 = 1$, user 2 operates in a time-division manner with users 1 and 3. As $a_1$ grows, both $\theta_1$ and $\theta_2$ increase and thus users 1 and 2 transmit in an overlapping fashion. Meanwhile, $\theta_3$ decreases such that users 2 and 3 still operate in a time-division manner, which is consistent with the results in Section \ref{subsec:twouserResults} for a two-user Gaussian ZIC with weak interference. In this regime, which takes place for $1\leq a_1\leq 2.1$, $\theta_3$ and thus the achievable rate of user 3 are sacrificed to get higher transmission fractions and rates for users 1 and 2. For larger $a_1$, $\theta_1$ reaches its optimal value in the single-user case, $\theta_1^* = 0.87$, and $\theta_2$ and $\theta_3$ remain constant at levels that are less than their counterparts in the single-user case. The observations above show the interdependency of $(\theta_1,\theta_2,\theta_3)$ in the CGZIC setting.

\begin{figure}
\centering
\includegraphics[width = 3.5in]{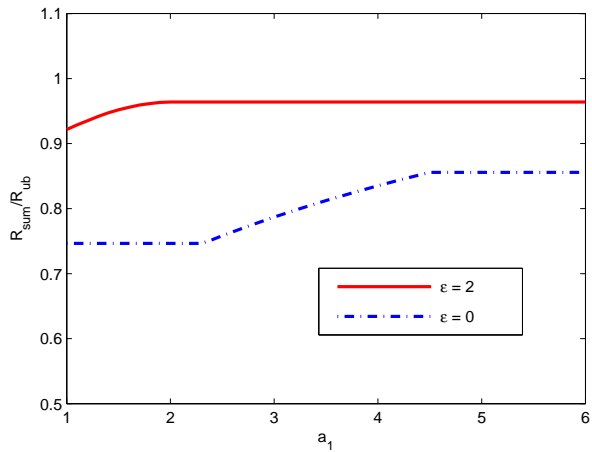}
\caption{Normalized maximum achievable sum rates for the three-user CGZIC with processing energy cost $\epsilon_1 = \epsilon_2 = \epsilon_3 = \epsilon$, as a function of cross-link power gain $a_1$ for $\epsilon = 2$ and $\epsilon = 0$ when $a_2 = 0.5$, $P_1 = 4$, $P_2 = 3.5$ and $P_3 = 3$, where $R_{sum} = R_{sum,IV}$ and $R_{ub}$ is the interference-free upper bound.}
\label{fig:CGZICRatio}
\end{figure}

Finally, as in Fig. \ref{fig:ICRatio} of Section \ref{subsec:twouserResults}, in Fig. \ref{fig:CGZICRatio}, we compare the maximum achievable rate $R_{sum}$ (in this case, that of Scheme IV), normalized by the interference-free upper bound $R_{ub}$, i.e., the ratio $R_{sum}/R_{ub}$, for the cases with and without processing energy cost in the cascade Z set-up. Similar improvements on the normalized sum rate are observed here as well, showing that for the three-user CGZIC, leveraging bursty transmission and scheduling user transmission bursts appropriately based on the channel set-up are essential in mitigating the effect of interference in the presence of the processing energy cost.

\subsection{Discussion: Other Regimes for $(a_1,a_2)$}
The achievable schemes studied in the previous subsections apply to the three-user CGZIC with processing energy cost in the mixed regime $a_1\geq 1$ and $0<a_2<1$. For general power gains $(a_1,a_2)$, Scheme I, II and III follow in a similar manner. Below, we briefly discuss how Scheme IV would be modified for different ranges of $(a_1,a_2)$. For $a_1\geq 1$ and $a_2\geq 1$, in Scheme IV, both receivers 2 and 3 would benefit from overhearing their respective interference signals when their own transmitters are silent, leading to a condition similar to (\ref{eqn:burJointRate1}) for $R_2$. When $0<a_1<1$ and $a_2\geq 1$, in Scheme IV, only receiver 3 would benefit from overhearing the interference when its own transmitter is off. Finally, for $0<a_1\leq 1$ and $0<a_2\leq 1$, Scheme IV would not be applicable, since in this regime, either of the interfered receivers cannot cancel the interference through overhearing of the interference signals.

\section{Conclusions}
In this paper, we have studied the impact of transmit processing energy cost on the achievable sum rate of Gaussian interference channels. When the processing energy cost is present, it is no longer optimal for each transmitter to transmit all the time as in the conventional no processing cost case. For the two-user Gaussian IC and the three-user CGZIC, we have proposed transmission schemes with relatively low complexities for the purpose of maximizing the sum rate. The results from the former model highlight how bursty transmission due to processing energy cost can be leveraged to mitigate the effect of interference. We have also found that, with processing energy cost, a larger range of cross-link power gains could ensure the very strong interference condition compared with its counterpart in the no processing cost case.  The investigation of the latter model suggests that one should take into consideration the channel set-up when scheduling user transmissions. Future work and directions include study of more practical but tractable modeling of the processing energy cost and extensions to fading scenarios.

\bibliographystyle{IEEEtran}
\bibliography{IEEEabrv,processingcost2}

\begin{thebibliography}{10}
\providecommand{\url}[1]{#1}
\csname url@samestyle\endcsname
\providecommand{\newblock}{\relax}
\providecommand{\bibinfo}[2]{#2}
\providecommand{\BIBentrySTDinterwordspacing}{\spaceskip=0pt\relax}
\providecommand{\BIBentryALTinterwordstretchfactor}{4}
\providecommand{\BIBentryALTinterwordspacing}{\spaceskip=\fontdimen2\font plus
\BIBentryALTinterwordstretchfactor\fontdimen3\font minus
  \fontdimen4\font\relax}
\providecommand{\BIBforeignlanguage}[2]{{%
\expandafter\ifx\csname l@#1\endcsname\relax
\typeout{** WARNING: IEEEtran.bst: No hyphenation pattern has been}%
\typeout{** loaded for the language `#1'. Using the pattern for}%
\typeout{** the default language instead.}%
\else
\language=\csname l@#1\endcsname
\fi
#2}}
\providecommand{\BIBdecl}{\relax}
\BIBdecl

\bibitem{processingcost:Liu_Erkip_Asilomar11}
X.~Liu and E.~Erkip, ``On the {G}aussian {Z} interference channel with
  processing energy cost,'' in \emph{Proceedings of Asilomar Conference on
  Signals, Systems and Computers}, Pacific Grove, CA, Nov. 2011.

\bibitem{processingcost:Massaad_Medard_Zheng04}
P.~{Youssef-Massaad}, M.~Medard, and L.~Zheng, ``Impact of processing energy on
  the capacity of wireless channels,'' in \emph{Proceedings of IEEE ISITA},
  Oct. 2004.

\bibitem{processingcost:Massaad_Medard_Zheng_TWC08}
------, ``Bursty transmission and glue pouring: on wireless channels with
  overhead costs,'' \emph{IEEE Trans. Wireless Commu.}, vol.~7, no.~12, pp.
  5188--5194, 2008.

\bibitem{processingcost:Massaad_Medard_Zheng_Allerton}
------, ``On the effect of processing energy on multiple access channels,'' in
  \emph{Proceedings of Allerton Conference}, Monticello, IL, Oct. 2004.

\bibitem{processingcost:Kramer04}
G.~Kramer, ``Models and theory for relay channels with receive constraints,''
  in \emph{Proceedings of Allerton Conference}, Monticello, IL, Oct. 2004.

\bibitem{references:ElGamal_Kim_nit11}
A.~E. Gamal and Y.~H. Kim, \emph{Network Information Theory}.\hskip 1em plus
  0.5em minus 0.4em\relax Cambridge University Press, 2011.

\bibitem{processingcost:Yuanpeng_Erkip_isit11}
Y.~Liu and E.~Erkip, ``On the sum capacity of {K}-user cascade {Gaussian}
  {Z}-interference channel,'' in \emph{Proceedings of IEEE ISIT}, Saint
  Petersburg, Russia, Jul. 2011.

\bibitem{processingcost:Sato81}
H.~Sato, ``The capacity of the {G}aussian interference channel under strong
  interference,'' \emph{{IEEE} Trans. Inf. Theory}, vol.~27, pp. 786--788, Nov.
  1981.

\bibitem{processingcost:Han_Kobayashi81}
T.~S. Han and K.~Kobayashi, ``A new achievable rate region for the interference
  channel,'' \emph{{IEEE} Trans. Inf. Theory}, vol.~27, no.~1, pp. 49--60,
  1981.

\bibitem{processingcost:Shang09}
X.~Shang, G.~Kramer, and B.~Chen, ``A new outer bound and the
  noisy-interference sum-rate capacity for {G}aussian interference channels,''
  \emph{{IEEE} Trans. Inf. Theory}, vol.~55, no.~2, pp. 689--699, 2009.

\bibitem{processingcost:Zhang_Guo_isit11}
L.~Zhang and D.~Guo, ``Capacity of {G}aussian channels with duty cycle and
  power constraints,'' in \emph{Proceedings of IEEE ISIT}, Saint Petersburg,
  Russia, Jul. 2011.

\bibitem{processingcost:Sinha_Chandrakasan01}
A.~Sinha and A.~Chandrakasan, ``Dynamic power management in wireless sensor
  networks,'' \emph{IEEE Design {\&} Test of Computers}, vol.~18, no.~2, pp.
  62--74, 2001.

\bibitem{references:Khandani09}
A.~S. Motahari and A.~K. Khandani, ``Capacity bounds for the {G}aussian
  interference channel,'' \emph{{IEEE} Trans. Inf. Theory}, vol.~55, no.~2, pp.
  620--643, 2009.

\bibitem{processingcost:Etkin08}
R.~Etkin, D.~Tse, and H.~Wang, ``Gaussian interference channel capacity to
  within one bit,'' \emph{{IEEE} Trans. Inf. Theory}, vol.~54, no.~12, pp.
  5534--5562, 2008.

\bibitem{processingcost:Sason_IT04}
I.~Sason, ``On achievable rate regions for the {G}aussian interference
  channel,'' \emph{IEEE Trans. Inf. Theory}, vol.~50, no.~6, pp. 1345--1356,
  2004.

\end{thebibliography}

\end{document}